\documentclass[12pt,preprint]{aastex}
\usepackage{amssymb, amsmath, apjfonts, emulateapj5, natbib, color}
\input psfig.tex

\def\apj{{ApJ}}
\def\apjs{{ApJS}}

\def\hb{H$\beta$}

\def\lax{{$\mathrel{\hbox{\rlap{\hbox{\lower4pt\hbox{$\sim$}}}\hbox{$<$}}}$}}
\def\gax{{$\mathrel{\hbox{\rlap{\hbox{\lower4pt\hbox{$\sim$}}}\hbox{$>$}}}$}}
\def\simlt{\lower.5ex\hbox{$\; \buildrel < \over \sim \;$}}
\def\simgt{\lower.5ex\hbox{$\; \buildrel > \over \sim \;$}}

\def\mnras{{MNRAS}}

\def\pasp{{PASP}}

\def\percm2{cm$^{-2}$}

\def\solmass{$M_\odot$}

\def\msig{{$M_{\rm BH}-\sigma_*$}}

\slugcomment{To appear in {\it The Astrophysical Journal}.}
\shorttitle{Black Hole Virial Mass Estimator}
\shortauthors{HO \& KIM}

\begin{document}

\title{A Revised Calibration of the Virial Mass Estimator for Black Holes in Active Galaxies Based on Single-epoch H$\beta$ Spectra}

\author{Luis C. Ho\altaffilmark{1,2} and Minjin Kim\altaffilmark{3,4}}

\altaffiltext{1}{Kavli Institute for Astronomy and Astrophysics, Peking 
University, Beijing 100871, China}

\altaffiltext{2}{Department of Astronomy, School of Physics, Peking University, Beijing 100871, China}


\altaffiltext{3}{Korea Astronomy and Space Science Institute, Daejeon 305-348, 
Republic of Korea}

\altaffiltext{4}{University of Science and Technology, Daejeon 305-350, 
Republic of Korea}
\begin{abstract}
The masses of supermassive black holes in broad-line active galactic nuclei 
(AGNs) can be measured through reverberation mapping, but this method 
currently cannot be applied to very large samples or to high-redshift AGNs. 
As a practical alternative, one can devise empirical scaling relations, based 
on the correlation between broad-line region size and AGN luminosity and the 
relation between black hole mass and bulge stellar velocity dispersion, to 
estimate the virial masses of black holes from single-epoch spectroscopy.  We 
present a revised calibration of the black hole mass estimator for the commonly 
used H$\beta$ emission line.  Our new calibration takes into account the recent 
determination of the virial coefficient for pseudo and classical bulges.
\end{abstract}


\section{Introduction}

Supermassive black holes (BHs) play a fundamental role in many aspects of 
contemporary extragalactic astronomy.  However, direct methods to measure BH 
masses currently are limited to very nearby, largely inactive galaxies.  
Reverberation mapping (RM; Blandford \& McKee 1982; Peterson 1993) of 
broad-line active galactic nuclei (AGNs) provides a means of measuring the 
size ($R$) of their broad-line region (BLR), which, in combination with the 
velocity widths ($\Delta V$) of the broad emission lines, yields a virial 
estimate of the BH mass 

\begin{equation}
M_{\rm BH}({\rm RM}) = f\frac{R(\Delta V)^2}{G}.  
\end{equation}

\noindent
The virial coefficient $f$ depends on the kinematics, geometry, and 
inclination of the BLR.  In practice, it is set by normalizing the RM AGNs to 
the correlation between BH mass and bulge stellar velocity dispersion 
($M_{\rm BH}-\sigma_*$ relation; Ferrarese \& Merritt 2000; Gebhardt et al. 
2000; see Kormendy \& Ho 2013 for a review) established by local galaxies with 
direct BH mass measurements (e.g., Onken et al. 2004; Woo et al.  2010).  
Despite its utility, RM requires time-consuming, long-term spectroscopic 
monitoring, and, to date, only $\sim 50$, mostly nearby ($z$ \lax 0.1) AGNs 
have been studied in this manner (e.g., Peterson et al 2004; Barth et al. 2011; 
Grier et al. 2013).  To access more 
luminous quasars and especially higher redshift systems, an alternative, much 
more expedient mass estimator can be devised from single-epoch spectroscopy by 
taking advantage of the empirical correlation between BLR size and continuum 
luminosity, $R \propto L^\gamma$ (Kaspi et al. 2000), to estimate $R$.  Then, 
from a straightforward measure of $L$ and $\Delta V$ from single-epoch or mean 
(time-averaged) spectra of the RM sample, one can solve for 

\begin{equation}
\log M_{\rm BH}({\rm RM}) = \log {\rm VP(mean)} + a,
\end{equation}

\noindent
where $a$ is a zero point offset and we define the virial product as

\begin{equation}
{\rm VP(mean)} = 
\left(\frac{\Delta V}{1000\,{\rm km\,s^{-1}}} \right)^2
\left(\frac{\lambda L_{\lambda}}{10^{44}\,{\rm erg\,s^{-1}}} \right)^\gamma.
\end{equation}

\noindent
The line width $\Delta V$ can be parameterized as the full-width at half 
maximum (FWHM) or line dispersion ($\sigma_{\rm line}$, second moment) of the 
broad line profile.  The above approach to calibrate ``single-epoch''
spectra has been developed by a number of authors (e.g., McLure \& Dunlop 2001; McLure \& Jarvis 2002; Vestergaard 2002; Vestergaard \& Peterson 2006; Wang et al. 2009).  For lower redshift sources ($z$ \lax 0.75), the most commonly used emission line is H$\beta$, and the continuum luminosity is referenced to 5100 \AA.

Some recent developments justify a reassessment of the virial formalism to 
estimate BH masses from single-epoch \hb\ spectroscopy.  Kormendy \& Ho 
(2013) significantly updated the \msig\ relation for inactive galaxies, 
highlighting, in particular, the large and systematic differences between the 
relations for pseudo and classical bulges.  This prompted Ho \& Kim (2014) to 
calibrate the $f$ factor separately for the two bulge types, using the latest 
sample of RM AGNs for which reliable bulge classifications could be performed. 
The changes are not negligible. Whereas previous studies obtained $f \approx 4.2-5.5$ (Onken et al. 2004; Woo et al. 2010; Park et al. 2012; Grier et al. 2013), Ho \& Kim find $f=6.3\pm1.5$ for classical bulges and ellipticals and $f = 3.2\pm0.7$ for pseudobulges.  (For the present discussion, we refer to $f$ calculated using $\Delta V = \sigma_{\rm line}$.)  Apart from these revisions to the \msig\ relation and the new calibration of the $f$ factor for pseudo and classical bulges, the $R-L$ relation itself has been updated by Bentz et al. (2013).

\section{Calibration}

Our new calibration of Equation~2 uses the updated database of properties for 
the RM AGNs and the bulge type classifications of their host galaxies given in 
Ho \& Kim (2014), as summarized in Table~1.  Specifically, we compute 
$M_{\rm BH}({\rm RM}) = f {\rm VP}(\sigma_{\rm line})$, using $f$ and 
${\rm VP}(\sigma_{\rm line})$ derived from root-mean-square (rms) spectra.  We 
adopt $f=6.3\pm1.5$ for classical bulges 

\vskip 0.3cm
\begin{figure*}[t]
\centerline{\psfig{file=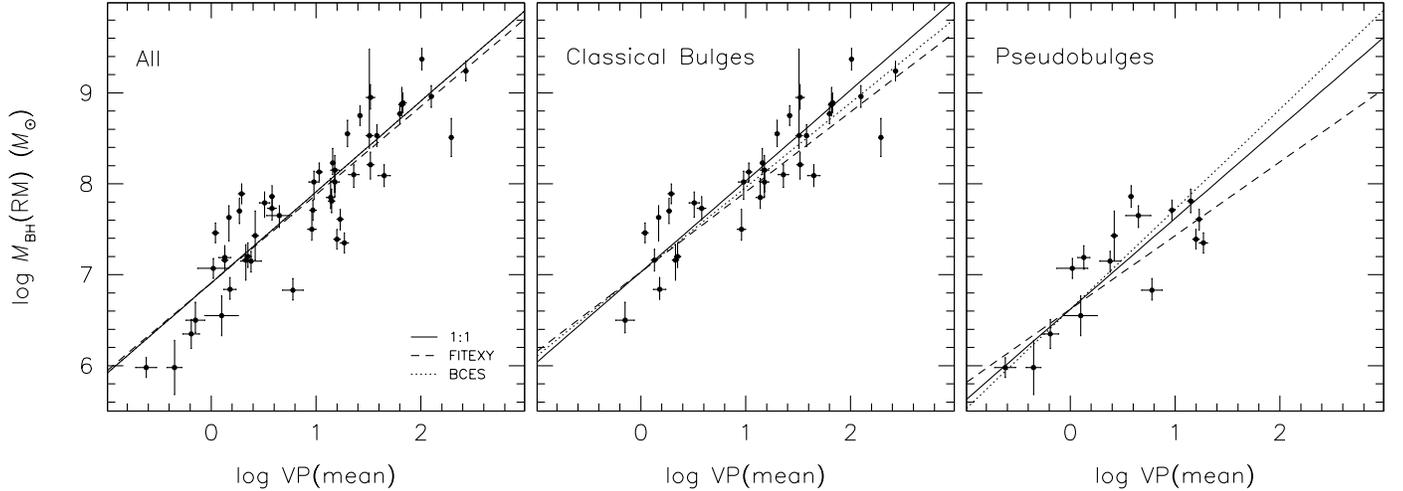,width=17.5cm,angle=270}}
\figcaption[fig1.ps]{Comparison between RM BH masses and virial products
calculated from FWHM of \hb\ extracted from mean spectra for (left) the
entire sample, (middle) classical bulges, and (right) pseudobulges.  The
solid line denotes a slope of unity, while the dashed and dotted lines
give the regression based on the {\tt FITEXY} and {\tt BCES} fits,
respectively.
\label{fig1}}
\end{figure*}
\vskip 0.3cm

\noindent
and ellipticals and $f = 3.2\pm0.7$ for pseudobulges.  As an 
approximation to quantities pertaining to single-epoch spectra, we use the 
values of $\lambda L_{\lambda}$(5100 \AA) and FWHM(\hb) derived from mean 
spectra, as listed in Table~2 of Ho \& Kim (2014).  Of the 43 RM AGNs that 
have bulge type classifications, 38 have available \hb\ line widths extracted 
from mean spectra, among them 14 pseudo and 24 classical bulges.  Among 
the empirical several criteria recommended by Kormendy \& Ho (2013; see their 
Supplemental Material) to distinguish pseudobulges from classical bulges, we 
adopt, whenever possible, the most widely used condition: S\'ersic (1968) 
index $n < 2$. However, the S\'ersic index of the bulge can be difficult to 
measure accurately in the presence of a bright nucleus. Under these 
circumstances, Ho \& Kim (2014) use a surrogate, equally effective 
criterion, that the bulge-to-total light fraction should be \lax 1/3.
Six objects have more than one set of observations; in our analysis we 
treat these multiple data points as independent measurements.  We have 
verified that combining the multiple measurements as a weighted average does 
not alter our final conclusions.


Figure~1 confirms that $M_{\rm BH}({\rm RM})$ correlates tightly and linearly 
with VP(mean).  A linear regression using the {\tt FITEXY} estimator (Press 
et al. 1992, as modified by Tremaine et al. 2002) yields a formal slope of 
$0.97\pm0.07$ and an intrinsic scatter of 0.35 dex for the entire sample with 
both bulge types combined. The {\tt BCES} algorithm of Akritas \& Bershady 
(1996) gives a consistent slope of $1.00\pm0.10$.  The middle and right panels 
of the figure show the correlations for the two bulge types separately. 
Although the {\tt FITEXY} fit seems to suggest that pseudobulges have a slope
less than unity, this is not supported by the {\tt BCES} fit.  Overall there 
is no evidence that the two bulge types behave differently.

Adopting a slope of $\gamma = 0.533^{+0.035}_{-0.033}$ for the latest $R-L$ 
relation (Bentz et al. 2013), the final mass scaling relation becomes

\begin{equation}
\log M_{\rm BH}({\rm H}\beta) = \log
\left[
\left(\frac{{\rm FWHM({\rm H}\beta}}{1000\,{\rm km\,s^{-1}}} \right)^2 
\left(\frac{\lambda L_{\lambda}({\rm 5100\,\AA})}{10^{44}\,{\rm erg\,s^{-1}}} \right)^{0.533} \right] + a,
\end{equation}

\noindent
where $a= 7.03\pm0.02$ for classical bulges and $a = 6.62\pm0.04$ for 
pseudobulges, with a corresponding intrinsic scatter of 0.32 and 0.38 dex. 
For both bulge types combined, $a = 6.91\pm0.02$ with an intrinsic scatter of 
0.35 dex, essentially identical to the results of Vestergaard \& Peterson 
(2006).  It is remarkable that our new fit, which is based on a larger 
sample and significant updates to all of the RM and velocity dispersion data, 
turns out to be so similar to that of Vestergaard \& Peterson published almost 
a decade ago.  This may indicate that this type of calibration is still 
currently dominated by systematic effects (e.g., intrinsic differences in the 
factor $f$).

\section{Concluding Remarks}

We present a new calibration of the prescription for estimating BH masses for 
broad-line AGNs and quasars using single-epoch spectra of the \hb\ emission 
line.  The primary difference between this and the previous work of 
Vestergaard \& Peterson (2006) is that we account for the systematic 
difference in the virial coefficient $f$ between pseudo and classical bulges 
(Ho \& Kim 2014).  We explicitly assume that AGNs hosted by pseudobulges 
follow a different \msig\ relation than those hosted by classical bulges, as 
observed in inactive galaxies.
In addition, our calibration sample of RM AGNs is 
significantly larger and more current than that used by Vestergaard \& 
Peterson, which was based on the sample of Peterson et al. (2004) from a 
decade ago.  Incorporating also recent minor changes to the $R-L$ relation, we 
find that the zero point of the \hb\ virial mass formalism for classical 
bulges is a factor of 2.6 (0.41 dex) higher than that for pseudobulges.  The 
commonly used \hb\ calibration of Vestergaard \& Peterson (2006) is very 
similar to ours for classical bulges, but their zero point is a factor of 2 
higher than our zero point for pseudobulges.

This important source of systematic uncertainty obviously should be eliminated 
to the extent possible.  However, as discussed in Ho \& Kim (2014), in practice
this will prove challenging because of the difficulty of measuring accurate 
host galaxy parameters for luminous, and especially distant, AGNs.  As a 
general rule of thumb, any system with $M_{\rm BH}$ \gax\ $10^8$ \solmass\ 
can be safely regarded as a classical bulge or elliptical galaxy.  This 
can be seen from the distribution of BH masses in inactive galaxies (Ho \& Kim 
2014, Fig. 2), as well as for the RM-mapped AGNs in this paper (right panel of 
Fig. 1).  At the other extreme, BHs with $M_{\rm BH}$ \lax\ $10^6$ \solmass\ 
almost certainly reside in pseudobulges (e.g., Greene et al. 2008; Jiang et al. 
2011).  Between these two limiting cases ($M_{\rm BH} \approx 10^6 - 10^8$ 
\solmass), BH masses estimated from single-epoch spectroscopy presently 
cannot be known to better than a factor of $\sim 2$ without knowledge of the 
bulge type of the host galaxy.

\acknowledgements
We thank the referee for valuable criticisms.  LCH acknowledges support by the Chinese Academy of Science through grant No. XDB09030102 (Emergence of Cosmological Structures) from the Strategic Priority Research Program and by the National Natural Science Foundation of China through grant No. 11473002.






\clearpage
\begin{figure*}[t]
\centerline{\psfig{file=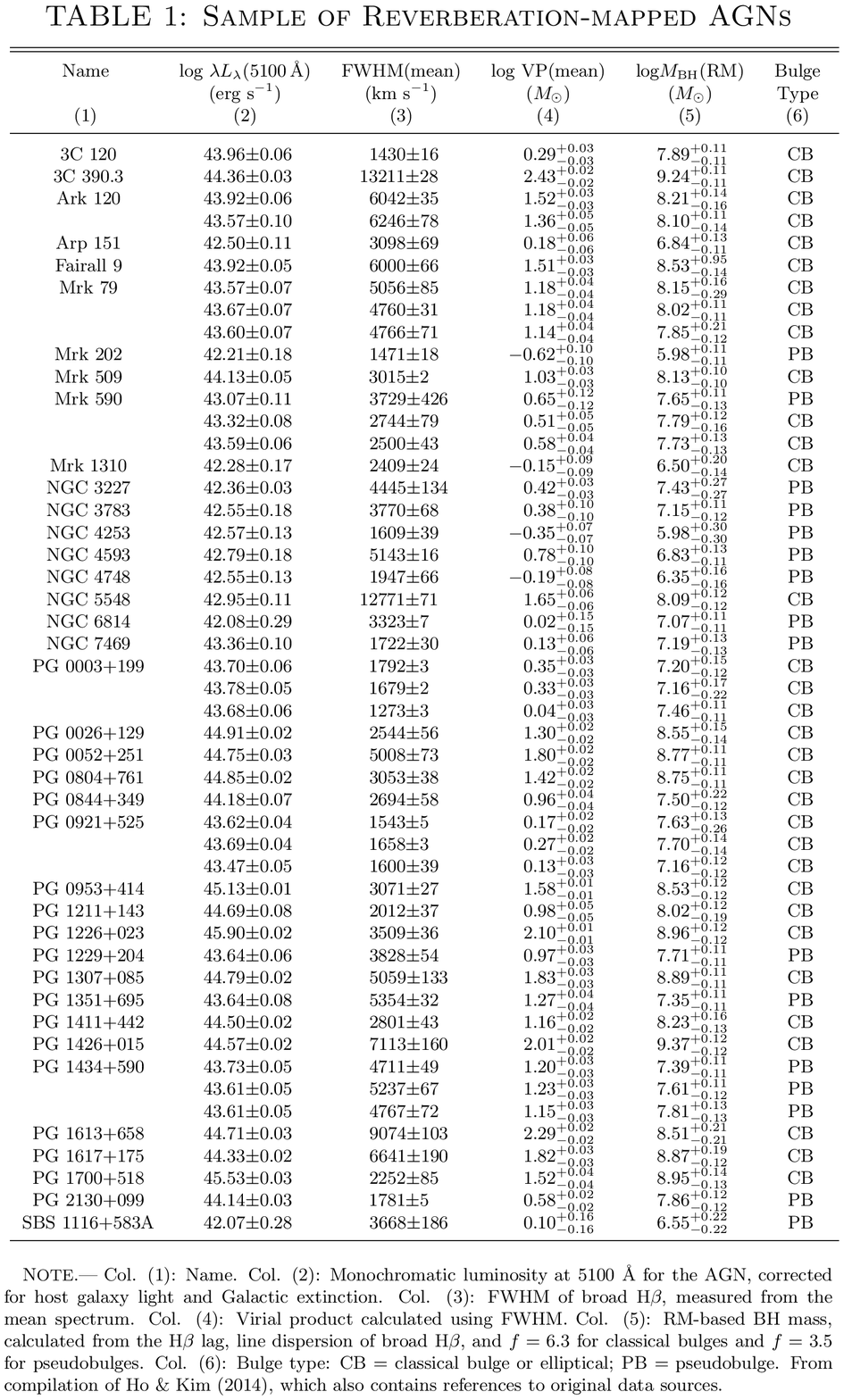,width=7.3in,angle=0}}
\end{figure*}
\end{document}